\documentclass{ws-ijmpd}
\usepackage[super,compress]{cite}
\begin{document}

%\markboth{} {}

\catchline{}{}{}{}{}
%
%%%%%%%%%%%%%%%%%%%%%%%%%%%%%%%%%%%%%%%%%%%%%%%%%%%%%%%%%%%%%%%%%%%%

\title{PARTICLES ON THE ROTATING CHANNELS IN THE WORMHOLE METRICS
%\footnote{For the title, try not to use more than 3
%lines. Typeset the title in 10~pt Times roman, uppercase and
%boldface.}
}

\author{G. ARSENADZE
}

\address{School of Physics, Free University of Tbilisi\\
Tbilisi, 0159, Georgia\\
garse15@freeuni.edu.ge}

\author{Z. OSMANOV}

\address{School of Physics, Free University of Tbilisi\\
Tbilisi, 0159, Georgia\\
z.osmanov@freeuni.edu.ge
}

\maketitle

\begin{history}
\received{Day Month Year}
\revised{Day Month Year}
\end{history}

\begin{abstract}
In the Ellis-Bronnikov wormhole (WH) metrics the motion of a particle along curved rotating channels is studied. By taking into account a prescribed shape of a trajectory we derive the reduced $1+1$ metrics, obtain the corresponding Langrangian of a free particle and analytically and numerically solve the corresponding equations of motion. We have shown that if the channels are twisted and lag behind rotation, under certain conditions beads might asymptotically reach infinity, leaving the WH, which is not possible for straight co-rotating trajectories. The analytical and numerical study is performed for two and three dimensional cases and initial conditions of particles are analysed in the context of possibility of passing through the WH. 
\end{abstract}


\ccode{PACS numbers: Need to be added!}

%%%%%%%%%%%%%%%%%%%%%%%%%%%%%%%%%%%%%%%%%%%%%%%%%%%%%%%%%%%%%%%%%%%%%
\section{Introduction}
%
%
%
%
%%%%%%%%%%%%%%%%%%%%%%%%%%%%%%%%%%%%%%%%%%%%%%%%%%%%%%%%%%%%%%%%%%%%%
 
 In the scientific literature the possibility of existence of wormholes (WH) is actively discussed. It is believed that they might have been created in the early evolution of the universe (Refs. \refcite{wheeler}-\refcite{hawking}). It is worth noting that some authors consider the existence of accretion disks inside the WH metrics (Ref. \refcite{accret}). Therefore, accreting matter, presumably plasma, will rotate and if loaded by magnetic field, typical for accretion disks, of the order of $10-100$ G (Ref. \refcite{AGN}), the particles might be in frozen in condition following "field lines". Certain massive astrophysical objects may be entrances to WHs and according to a theoretical analysis these objects might have extremely strong magnetic fields of the order of $10^{13}$G or even higher (Ref. \refcite{kns}). 
 
In such an enormous value of magnetic field the charged particles will be strongly bound by the "field lines". One can straightforwardly show that gyroradius of protons, $\gamma m_pc^2/eB$, where $\gamma$ is the Lorentz factor of the particle, $m_p$ and $e$ are its mass and charge respectively and $B$ is the magnetic induction, is of the order of $1$cm for ultrarelativistic protons with $\gamma\sim 10^7$. Much less will be the corresponding radius for electrons. Therefore the particles will follow the magnetic field lines, sliding along them. Particularly interesting scenarious could potentially happen if these field lines rotate and the particles experience relativistic centrifugal force.
 
A very simplified model has been considered in (Ref. \refcite{mr}) where the authors studied dynamics of particles sliding along a rectilinear rotating channels and found an unusual behaviour of relativistic centrifugal force. It has been shown that reaching the light cylinder (LC) zone (area where the linear velocity of rotation exactly coincides with the speed of light) the radial acceleration will inevitably change its sign and under certain conditions it always might be negative. It is believed that this is a mechanism alternative to the Fermi-type acceleration (Ref. \refcite{fermi}) and in certain cases even more efficient (Ref. \refcite{rieger}). In particular, there is a series of papers dedicated to the study of astrophysical application of relativistic centrifugal effects (Refs. \refcite{applic1}) and modified magnetocentrifugal acceleration in the magnetospheres of pulsars, supermassive and stellar mass black holes (Refs. \refcite{applic2}). 

In the mentioned publications particle dynamics has been studied for rectilinear magnetic field lines. But in real astrophysical situations the field lines are curvilinear, especially close to the LC, where for not violating the relativity principle the field lines twist, lagging behind the rotation.  On the other hand, observations on cosmic rays indicate that, since they have reached us they should be in force-free regime, which means that the field lines might have shapes of the Archiimede's spiral. The corresponding mechanism of reconstruction of magnetospheres in the LC area has been studied for pulsars and supermassive black holes (Ref.  \refcite{reconstr}) and it has been shown that under certain circumstances the field lines obtain the configuration of the Archimede's spiral.

Acceleration of particles moving along prescribed rotating Archimede's spiral has been considered in (Ref. \refcite{rdo}), where it was shown that the mentioned configuration enables particles to leave the LC zone. Corresponding three dimensional generalisation has been studied for the Crab pulsar (Ref. \refcite{gud}). In the present work we apply the developed model to study the dynamics of particles on rotating magnetic field lines in the WH metrics. This is a first attempt of this kind, therefore, as an initial step, we examine the most simplified Ellis-Bronnikov WH metrics (see Refs. \refcite{elli}-\refcite{kip}) and consider the dynamics of particles to understand conditions when beads freely leave the inner region of the WH. 

The paper is organised in the following way. In section~II, a theoretical model of dynamics of particles sliding along prescribed rotating channels in the WH metrics is presented, in section~III we consider results and in section ~IV we summarise them emphasizing the most significant results.

%%%%%%%%%%%%%%%%%%%%%%%%%%%%%%%%%%%%%%%%%%%%%%%%%%%%%%%%%%%%%%%%%%%%%
\section{Main consideration} \label{sec:main}
As it has already been mentioned, we examine the simplified Ellis-Bronnikov Wormhole metrics which can be written as (see Refs. \refcite{elli}-\refcite{kip}):
\begin{equation}
ds^2=-d t^2+dl^2+(b_0^2+l^2)(d\theta^2+\sin^2\theta d\phi^2)
\end{equation}
where $t$ is a time coordinate, $\theta$ and $\phi$ are spherical coordinates, $l$ is a coordinate measuring proper radial distance at fixed $t$, $b_0$ is constant and can be imagined as effective radius of Wormhole's throat and throughout the paper we use units where $c=G=1$. 
We parametrise a curve in the following way:
\begin{equation}
\theta=\theta(l)
\end{equation}
and
\begin{equation}
\varphi=\varphi(l) .
\end{equation}
These two functions fully fix the field line and give us the opportunity to examine particles moving along the line. We are interested in the effects of rigid rotation imposed on the aforementioned metrics. Generally speaking, there is a class of WHs which are intrinsically characterised by rotation, but the corresponding metric becomes more complex. The aim of the present paper is to make the first attempt of studying dynamic of particles on rotating trajectories imbedded in the WH metrics, therefore, the latter is given by the simplest static form. In this case rotation might be introduced by a matter flowing through the throat of the WH having nonzero angular momentum. Without going into details we assume that the matter is magnetised that is a common situation in typical astrophysical situations (Ref. \refcite{accret}).

For describing dynamics of particles on corotating field lines, one has to consider the azimuthal coordinate, $\phi$, in the laboratory frame of reference
\begin{equation}\label{phitot}
\phi=\varphi(l)+\omega t
\end{equation}

Rewriting the Wormhole metrics leads to the following expression
\begin{equation}
ds^2=g_{00}dt^2+2g_{01}dldt+g_{11}dl^2,
\end{equation}
with the corresponding metric tensor
\begin{equation}
\
g_{\alpha\beta}=
\begin{bmatrix}\label{metrics}
    -1+\omega^2\sin^2\theta(l)\left(b_{0}^2+l^2\right) & \varphi'(l)\omega \sin^2\theta(l)\left(b_{0}^2+l^2\right) \\
    \varphi'(l)\omega \sin^2\theta(l)\left(b_{0}^2+l^2\right) & \left(1+\left(\theta'(l)^2+\sin^2\theta(l)\varphi'(l)\right)\left[b_{0}^2+l^2\right]\right)
\end{bmatrix}
\
\end{equation}
\begin{equation}
\alpha,\beta=\{0,1\}
\end{equation}
Dynamics of relativistic particles on the prescribed co-rotating channels are described by the equation of motion
\begin{equation}\label{eqmo}
\frac{\partial L}{\partial x^\alpha}= \frac{d}{d\tau} \left(\frac{\partial L}{\partial\dot{x}^\alpha}\right)
\end{equation}
with the Lagrangian
\begin{equation}
L=-\frac{1}{2} g_{\alpha \beta} \frac{dx^\alpha}{d\tau} \frac{dx^\beta}{d\tau}
\end{equation}
where in the framework of the paper we use the following notations:  $x^0=t$, $x^1=l$ and $\dot{x}^\alpha={dx^\alpha}/{d\tau}$.

From Eq. (\ref{eqmo}) one finds
\begin{equation}\label{energy1}
\frac{\partial L}{\partial t}= \frac{d}{d\tau} \left(\frac{\partial L}{\partial\dot{t}}\right),
\end{equation}
but since the Lagrangian does not depend properly on time, the aforementioned equation leads to the conserved quantity of energy
\begin{equation}\label{energy2}
 E=-\gamma(g_{00}+g_{01}v)=const
\end{equation}
where $v={dl}/{dt}$ and
\begin{equation}\label{gamma}
\gamma=(-g_{00}-2g_{01}v-g_{11}v^2)^{-\frac{1}{2}}
\end{equation}
is the corresponding Lorentz factor. Solving the system of Eqs. (\ref{energy2},\ref{gamma}) versus $\upsilon$ one obtains (Ref. \refcite{gud})
\begin{equation}\label{v}
v=\frac{\sqrt{g_{00}+E^2}}{(g_{01}^2+E^2g_{11})}\left[-g_{01}\sqrt{g_{00}+E^2}\pm E\sqrt{g_{01}^2-g_{00}g_{11}} \right].
\end{equation}

%\begin{equation}
%v=-\frac{\omega}{\varphi'(l)} +\frac{\varphi'(l)\omega\sin^2\theta(1+E^2) \pm E\sqrt{(\omega^2\sin^2\theta+\frac{E^2-1}{r^2})(\sin^2\theta(\varphi'(l)-\omega^2)+\frac{1}{r^2}))}}{r^2(\varphi'(l)\omega^2\sin^4\theta+\frac{E^2}{r^4}+\frac{E^2}{r^2}\varphi'(l)sin^2\theta)}
%\end{equation}
%where
%\begin{equation}
%r=(b_0^2+l^2)
%\end{equation} 

%This equations shows that as $l\rightarrow \infty$ the velocity becomes constant.
%\begin{equation}
%v=-\frac{\omega}{\varphi'(l)}.
%\end{equation}

%Equation $(21)$ also shows what happens to velocity if $\theta=0$. If $\theta=0$ we get: 
%\begin{equation}
%v=\sqrt{1-\frac{1}{E^2}}
%\end{equation}

This expression shows behaviour of the radial velocity with the radial coordinate, $l$. On the other hand, it is worth noting that for studying dynamics of the particle one has to consider Lagrange's equations Eq. (\ref{eqmo}), which after straightforward calculations lead to the following expression for the radial acceleration
\begin{equation}\label{acc}
\frac{d^2l}{dt^2}=\frac{g_{00}+g_{01}v}{2g_{00}g_{11}-g_{01}^2}\left[A+\left(B+\frac{g_{01}}{g_{00}}C\right)v^2\right]+\frac{C}{g_{00}}v^3
\end{equation}
where we use the following notations
 \begin{equation}
A=2l\omega^2 \sin^2\theta(l) 
 \end{equation}
 \begin{equation}
 B=-2l\left[\theta'(l)^2+\varphi'(l)^2\sin^2\theta(l)\right]
 \end{equation}
 \begin{equation}
 C=\frac{2l\omega^2\sin^2\theta(l)}{v}+2l\varphi\omega\sin^2\theta(l)
 \end{equation}
%
%
%
%
%%%%%%%%%%%%%%%%%%%%%%%%%%%%%%%%%%%%%%%%%%%%%%%%%%%%%%%%%%%%%%%%%%%%%

%%%%%%%%%%%%%%%%%%%%%%%%%%%%%%%%%%%%%%%%%%%%%%%%%%%%%%%%%%%%%%%%%%%%%
\section{Discussion} \label{sec:res}
In this section we discuss two particularly interesting configurations of field lines. As a first example we examine channels having the form of the Archimedes' spiral in 2D. On the next step we examine a 3D case with $a\neq0$ and $\theta\neq \pi/2$. First of all we simplify equation Eq. (\ref{acc}) for $a=0$ and $\theta=const$ rewriting it in the following way
\begin{equation}\label{acc1}
\frac{d^2l}{dt^2}=\frac{\omega^2l\sin^2\theta}{1-(b_0^2+l^2)\omega^2\sin^2\theta}\left[1-(b_0^2+l^2)\omega^2\sin^2\theta-2v^2\right].
\end{equation}
As it is clear, it is very similar to an expression of the radial acceleration derived for the Minkowski space-time, except the terms $b_0$ and $\theta$, respectively being zero and $\pi/2$, in that case (see Ref. \refcite{mr}). One of the interesting features seen from the above equation is that if a particle initially is at the origin of the rotator, $l=0$, the radial acceleration will be negative from the very beginning if the initial radial velocity satisfies the condition 
\begin{equation}\label{vc}
v>\left(\frac{1-b_0^2\omega^2\sin^2\theta}{2}\right)^{1/2}.
\end{equation}
This is a generalisation of the result derived in the Minkowski space-time, which has been derived in (Ref. \refcite{mr}) and has been shown that the particle radially decelerates from the very beginning if the initial velocity exceeds $1/\sqrt{2}$.

If the particle moves from the inner region of the wormhole outwards, one can straightforwardly show from Eq. (\ref{v}) that by approaching the LC area the radial velocity behaving as
\begin{equation}\label{v1}
v\approx\left(1-\frac{b_0^2+l^2}{b_0^2+l_{LC}^2}\right)^{1/2},
\end{equation}
completely vanishes on the LC surface, where $l_{LC}$ is the light cylinder radius (see Eq. (\ref{metrics}))
\begin{equation}\label{lc}
l_{LC}=\left(\frac{1}{\omega^2\sin^2\theta}-b_0^2\right)^{1/2}.
\end{equation}
The aforementioned behaviour is a natural result, because in the mentioned location, the linear velocity of rotation exactly equals the speed of light and consequently the radial velocity must vanish. Substituting Eq. (\ref{v1}) into Eq. (\ref{acc1}) one can show that when a particle reaches the LC zone with the zero radial velocity, the corresponding acceleration becomes negative. This means that if the rigid rotation is maintained the particle never leaves the "interior" of the wormhole. This can easily be seen by introducing the effective potential, 
 \begin{equation}\label{potent1}
\frac{d^2l}{dt^2} = -\frac{\partial U}{\partial l}
\end{equation}
which after combining with Eq. (\ref{v}) leads to
 \begin{equation}\label{potent2}
U = \frac{l^2\omega^2\sin^2\theta}{2E^2}\left(E^2-2+2b_0^2\omega^2\sin^2\theta+l^2\omega^2\sin^2\theta\right).
\end{equation}

\begin{figure}
\resizebox{1.\textwidth}{!}{%
  \includegraphics{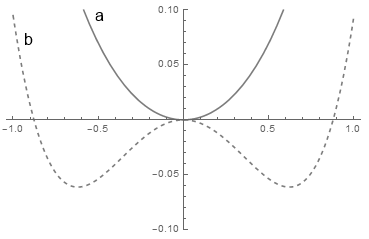}
    }
 \label{fig1}
    \caption{On this graph we present the effective potential of the particle as a function of the radial distance for two different values of energy. The set of parameters is following: $\omega = 1$, $\theta = \pi/2$, $b_0 = 0.1$, $l_0 = -5$, $E = 1.98$ (a) and $E = 1.1$ (b).}\end{figure}

In Fig.(1) we show the dependence of the effective potential on the radial coordinate for two different values of energy. The set of parameters is following: $\omega = 1$, $\theta = \pi/2$, $l_0 = -5$, $b_0 = 0.1$, $E = 1.98$ (a) and $E = 1.1$ (b). For the second case (b), the effective potential has more than one extremum, which means that particles initially accelerate, reach the point of the local minimum and proceed moving with radial deceleration, stopping at the LC surface. On the contrary, the case (a) corresponds to the initial radial velocity satisfying the condition presented by Eq. (\ref{vc}), therefore the particle from the very beginning radially decelerates and correspondingly the effective potential does not have the second local extremum.

On the other hand, it is clear that no real physical system can maintain rigid rotation. Close to the LC surface the particles become so energetic that they twist the magnetic field lines, which lag behind the rotation and as a result the plasma particles transit the LC area, avoiding the mentioned problem (see Refs. \refcite{reconstr}). It has been shown in these works that there is a region in the magnetosphere, where the field lines are twisted in a right direction, thus opposite to the rotation, allowing the particles leave the wormhole. Therefore, as a next and a more realistic example we consider curvilinear field lines.

\begin{figure}
\resizebox{1.\textwidth}{!}{%
  \includegraphics[height=5cm]{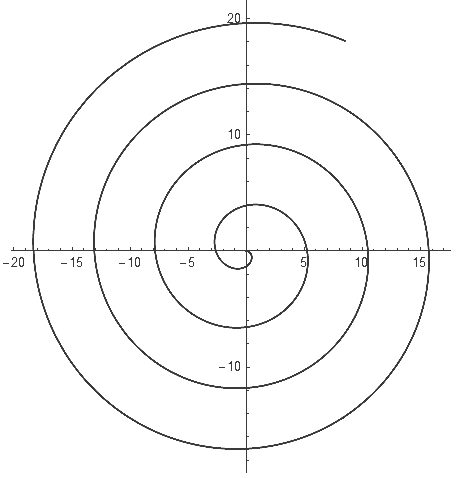}
    \includegraphics[height=5cm]{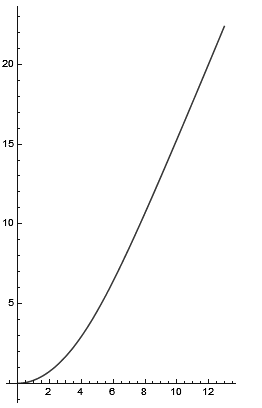}
    }
  \label{fig2}
  	 \caption{Trajectory of the particle in 2D as seen from the local frame of reference (left panel) and the laboratory frame of reference (right panel) respectively. The set of parameters is $\omega = 1$, $\theta = \pi/2$, $b_0 = 0.1$, $l_0 = 0$, $v_0 = 0.99$, $a=-1.2$.}\end{figure}
   
By taking into account the conditions, $\varphi=al$ and $\theta=const$, Eq. (\ref{acc}) reduces to
\begin{equation}  \label{acc3d}
\frac{d^2l}{dt^2}=-\frac{\Omega l\sin^2\theta\left[va-\omega+2v^2\omega+(b_0^2+l^2)\omega\Omega^2sin^2\theta\right]}{1+(b_0^2+l^2)(a^2-\omega^2)sin^2\theta},
\end{equation}
where $\Omega = \omega+av$ is the effective angular velocity of the particle as seen from the laboratory frame of reference (see Eq. (\ref{phitot})). From this equation it is evident that if the field lines are twisted such a way that $\Omega$ equals zero, the particles do not accelerate. The question is to understand how particles behave if their initial velocities do not coincide with $-\omega/a$. 

As a first example we examine dynamics on 2D curves located in the equatorial plane. In Fig. 2 we show the trajectory of the particle in the local frame of reference (left panel) and the laboratory frame of reference (right panel). The set of parameters is $\omega = 1$, $\theta = \pi/2$, $b_0 = 0.1$, $l_0 = 0$, $v_0 = 0.99$, $a=-1.2$. For this configuration of the trajectory, the initial velocity is different from the critical value, $-\omega/a\approx 0.83$, corresponding to the force free regime of dynamics. From the point of view of the local observer the particle moves along the spiral (left panel) but as we see from the right panel, from the laboratory frame of reference the particle's trajectory is drastically different from the Archimedes' spiral and in due course of time the shape tends to a straight line and consequently dynamics becomes force free.

\begin{figure}
\resizebox{1.\textwidth}{!}{%
  \includegraphics[height=4cm]{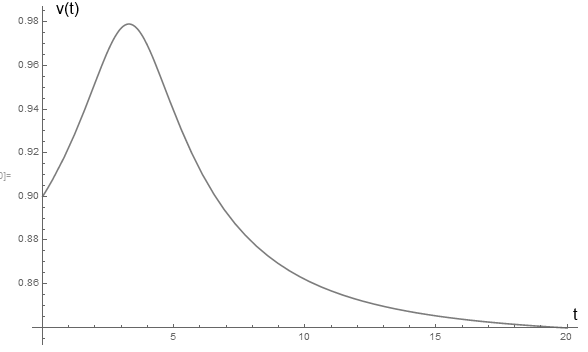}
    \includegraphics[height=4cm]{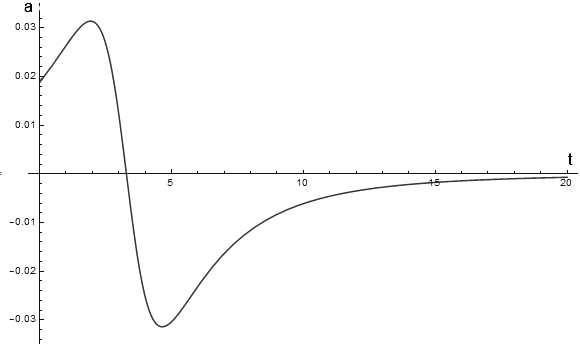}
    }
  \label{fig3}
   \caption{Trajectory of the particle in 2D as seen from the local frame of reference (left panel) and the laboratory frame of reference (right panel) respectively. The set of parameters is $\omega = 1$, $\theta = \pi/2$, $b_0 = 0.1$, $l_0 = 0$, $v_0 = 0.99$, $a=-1.2$.}\end{figure}

 In Fig. 3 we plot the graphs of the radial velocity (left panel) and radial acceleration (right panel) respectively. Despite the condition $v_0\neq -\omega/a$, in due course of time particles' velocity tends to the critical value (left panel) and acceleration asymptotically vanishes (right panel).

\begin{figure}
\resizebox{1.\textwidth}{!}{%
  \includegraphics{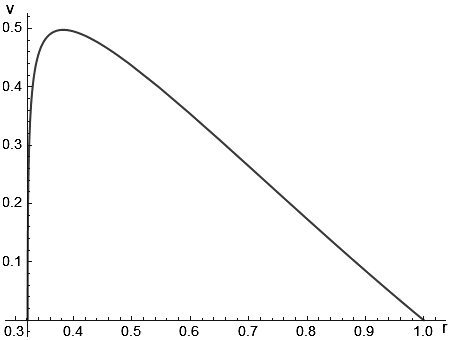}
    }
 \label{fig4}
    \caption{On this plot we present the behaviour of radial velocity versus the radial coordinate normalised by the LC radius (on the graph $r=\frac{l}{R_{LC}}$).The set of parameters is: $\omega = 1$, $\theta = \pi/2$, $b_0 = 0.1$, $l_0 = 3.1R_{LC}$, $v_0 = 0$, $a=-1.2$.}\end{figure}

In the mentioned example we have considered the archimedes' spiral with the critical velocity less than the speed of light, $-\omega/a<1$. As we have already discussed, the field lines might be less twisted in a sense that $-\omega/a>1$. In Fig. 4 we show the dependence of radial velocity on the normalised radial coordinate of a particle moving along a spiral with $-\omega/a=1/1.2$. The set of parameters is: $\omega = 1$, $\theta = \pi/2$, $b_0 = 0.1$, $l_0 = 3.1R_{LC}$, $v_0 = 0$, $a=-1.2$. For this shape of the field line there are no real particles moving with the radial velocities exceeding the speed of light, therefore, unlike the previous case, the condition $\Omega\rightarrow 0$ cannot be satisfied and asymptotically the angular velocity tends to $\omega$, the linear velocity of rotation coincides with the speed of light, the radial velocity vanishes and the particle stays inside the WH.

\begin{figure}
\resizebox{1.\textwidth}{!}{
  \includegraphics[height=5cm]{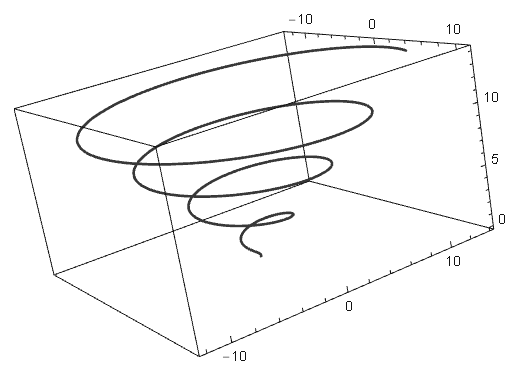}
    \includegraphics[height=5cm]{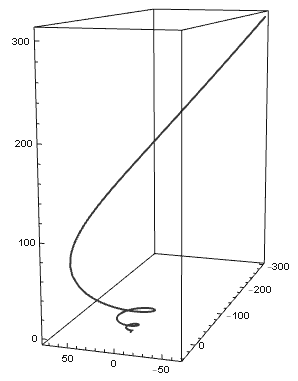}
    }
  \label{fig5}
   \caption{Trajectory of the particle in 3D as seen from the local frame of reference (left panel) and the laboratory frame of reference (right panel) respectively. The set of parameters is $\omega = 1$, $\theta = \pi/4$, $b_0 = 0.1$, $l_0 = 0$, $v_0 = 0.85$, $a=-1.2$.}\end{figure}

Considering the 3D generalisation is interesting not only from the theoretical point of view, but it is observationally evident that many active galactic nuclei reveal jet-like structures composed of relativistic particles. Therefore, it is significant to examine the configuration of 3D field lines. Since the mentioned particles are presumably in the force free regime, the field lines can be presented by $\varphi = al$ and $\theta=const$ (see Eq. (\ref{acc3d})).

In Fig. 5 we show the three dimensional analogues of Fig. 3. On the left panel we show the trajectory in the local frame of reference and on the right panel the trajectory in the laboratory frame of reference is shown. The set of parameters is $\omega = 1$, $\theta = \pi/4$, $b_0 = 0.1$, $l_0 = 0$, $v_0 = 0.85$, $a=-1.2$. As it is clear from these graphs by means of the special kind of shape of field lines in the local frame of reference, the trajectory although curved initially, asymptotically tends to a straight line. Such a behaviour can straightforwardly be shown from Eq. (\ref{v})
\begin{equation}
v=-\frac{\omega}{a} +\frac{1+E^2 \pm E\sqrt{a^2-\omega^2}}{\omega\sin^2\theta}\frac{1}{l^2},
\end{equation}
which confirms the fact that for asymptotically large values of $l$ the radial velocity achieves the critical value, $v_c$, and the effective angular velocity vanishes. This is demonstrated on Fig. 6 (see the left panel), where it is shown that despite different initial radial velocities of particles, all of them tend to $v_c$. On the right panel we plot the trajectories of particles as seen by a laboratory observer. As it is clear, particles sliding along rotating 3D Archimedes' spirals (see the left panel of Fig 5.) with appropriate initial conditions might leave the WH interior.

\begin{figure}
\resizebox{1.\textwidth}{!}{
  \includegraphics[height=4cm]{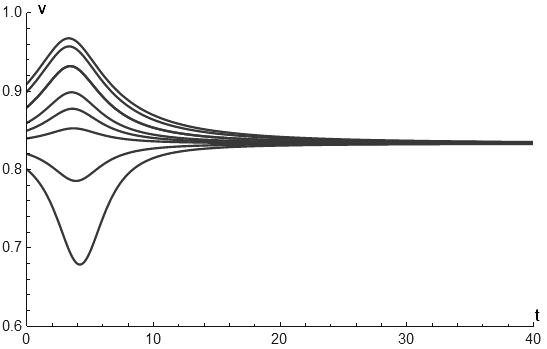}
    \includegraphics[height=4cm]{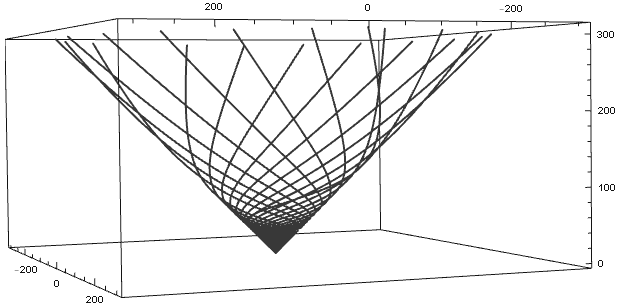}
    }
  \label{fig6}
   \caption{On the left panel we plot the behaviour of radial velocity for eight different initial radial velocities. On the right pane we show the trajectories of particles moving along the 20 similar trajectories with different initial angular shifts The set of parameters is $\omega = 1$, $\theta = \pi/4$, $b_0 = 0.1$,  $l_0 = 0$, $v_0 = 0.85$, $a=-1.2$.}\end{figure}
   
   \begin{figure}
\resizebox{1.\textwidth}{!}{%
  \includegraphics{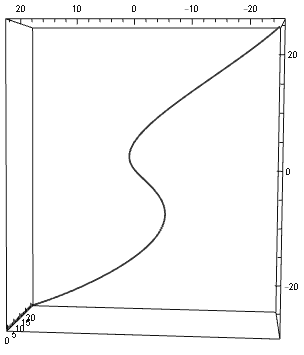}
    }
 \label{fig7}
    \caption{On this graph we plot the trajectory of a particle as observed by a laboratory observer. . The set of parameters is $\omega = 1$, $\theta = \pi/4$, $b_0 = 0.1$,  $l_0 = -40$, $v_0 = 0.95$, $a=-1.2$.}\end{figure}
   
  A very interesting regime is shown in Fig. 7 where we plot the trajectory of a particle observed in the laboratory frame of reference. The set of parameters is $\omega = 1$, $\theta = \pi/4$, $b_0 = 0.1$,  $l_0 = -40$, $v_0 = 0.95$, $a=-1.2$. As it is seen from the shape of the trajectory the particle starts moving relatively far from the WH with initial radial velocity different from $v_c=-\omega/a\approx 0.83$, then "enters" the inner region and leaves it.

 %%%%%%%%%%%%%%%%%%%%%%%%%%%%%%%%%%%%%%%%%%%%%%%%%%%%%%%%%%%%%%%%%%%%%
\section{Summary} \label{sec:summary}
%
%
%
%
%%%%%%%%%%%%%%%%%%%%%%%%%%%%%%%%%%%%%%%%%%%%%%%%%%%%%%%%%%%%%%%%%%%%%

\begin{enumerate}

     \item 
	We have developed a method for studying particle dynamics in rotating 2D and 3D trajectories imbedded in the simplest Ellis-Bronnikov WH metrics to understand conditions when the particles might potentially leave the interiors of WHs.
     \item For this purpose we have rewritten the metrics on a prescribed trajectories (field lines) and derived the equations of motion, governing dynamics of particles. 
      \item As a first example we have studied motion on straight rotating field lines. It has been shown that   depending on initial conditions the particles either radially accelerate or decelerate from the very beginning of motion. The process has been analysed also by means of the effective potential and it was found that in this case particles will never leave the WH.
      \item We have found that if dynamics is studied on the 2D Archimedes' spirals, under certain conditions the particles might asymptotically reach infinity, becoming force free. On the other hand, it has been shown that for a certain class of Archimedes' spirals, like the straight field lines, the particles will remain inside the WH interior.
       \item As a special and a more realistic example we have considered the 3D configuration of field lines having the properties of the flat Archimedes' spiral. For this particular case it has been shown analytically that dynamics of particles asymptotically tends to be force free and radial velocities of particles achieve a critical value if initially they were different from it. By numerically solving the equations of motion we have demonstrated how the trajectories would appear from the point of view of a laboratory observer and have described conditions when particles may potentially enter the WH interior and leave it.
     
 \end{enumerate}

The aim of this paper was to examine dynamics of the particle sliding along rotating magnetic field lines in the WH metrics to study necessary conditions when particles leave the interior of WHs. On the other hand, it is clear that in the framework of the paper we have considered a single particle approach, whereas it is significant to generalise this approximation to a more realistic situations and examine the motion of fluids inside the WH space-time. It is also interesting to study the similar problems for more complicated WH metrics. In the forthcoming papers we are going to consider the aforementioned problems and generalise the results presented in this work.
 
\section*{Acknowledgments}

The research was supported by the Shota Rustaveli
National Science Foundation grant (DI-2016-14).

\end{document}